\documentclass[]{ipart_v1}
\usepackage{amsbsy,enumerate}
\usepackage{graphicx}
\usepackage{graphics}

\firstpage{1}

\numberwithin{equation}{section}

\date{}
\usepackage[nodate]{datetime}

\begin{document}
\title[Unstable Behaviours of Classical Solutions...]{Unstable Behaviours of Classical Solutions in Spinor-type Conformal Invariant Fermionic Models}

\author[Fatma Aydogmus]{Fatma Aydogmus}

\begin{abstract}
It is well known that instantons are classical topological solutions existing in the context of quantum field theories that lie behind the standard model of particles. To provide a better understanding for the dynamical nature of spinor-type instanton solutions, conformal invariant pure spinor fermionic models that admit particle-like solutions for the derived classical field equations are studied in this work under cosine wave forcing. For this purpose the effects of external periodic forcing on two systems having different dimensions and quantum spinor numbers and have been obtained under the use of Heisenberg ansatz are investigated by constructing their Poincar\'e sections in phase space. As a result, bifurcations and chaos are observed depending on the excitation amplitude of the external forcing in both pure spinor fermionic models.
\end{abstract}

\maketitle

\section{Introduction}
The idea of describing bosons and fermions, having spin numbers up to 2 , by means of a fermionic field had been put forward by Louis de Broglie [1]. Moreover, Dirac$'$s nonlinear spinor field wave equation brought out a successful interpretation of electron and anti-electron (positron) system [2]. Later, Heisenberg and Pauli developed this equation in their unified field theory [3]. After these studies, great efforts were begun among theoretical physicists in searching new nonlinear massless conformal invariant field equations without dimensional parameters that especially would have classical solutions. In this context Gursey proposed a model as a possible basis for a unitary description of elementary particles which possesses broader dynamical symmetries compared to the Dirac$'$s and Heisenberg et al.$'$s works [4]. This model has been the first conformally invariant 4-D pure fermionic model and aroused a great interest. Later a class of its exact solutions with nonlinear $\left(\bar\psi\psi \right)^\frac{4}{3}$ self-coupled spinor term was found and they were shown to be instantonic in character [5, 6]. These classical solutions are similar to the solutions of pure Yang-Mills [7] equations in 4-D. 

Another spinor field system came with the work of Thirring [8] after Gursey’s one. Thirring$'$s conformally invariant model describes Dirac fermions in (1+1) space-time dimensions with no mass; but with non-linear self-interaction term  $\left(\bar\psi\psi \right)^2$  . It contains many typical features of the quantization of relativistic quantum field theories. Classical instanton-type solutions for this 2-D fermionic model were found too. It has been shown that these solutions were of the same form as the solutions of 4-D model [9].

In some cases, nonlinear classical field theories provide the possibility of deriving equation of motion of a particle as a singularity regarding from the point of field equation. But they have some disadvantages, like difficulties encountered in the solution and quantization of the equations. Some studies on the quantization and renormalization of the 4-D spinor wave equation have also been performed [10-13]. Recently, the role of the coupling constant in the evolution of 4-D spinor-type instantons has been studied [14]. After a while, dynamics of excited instantons in four-dimension have been investigated by the methods derived from nonlinear dynamical techniques [15]. Chaos has also been observed in the above mentioned 4-D dissipative nonlinear fermionic model [16]. 
In this paper, we re-consider the classical equations derived from the conformal invariant 2-D and 4-D pure spinor fermionic systems obtained by the Heisenberg$'$s ansatz. Firstly we compare the stability characterization of these spinor type instantons with that of Duffing equation. Secondly we periodically excited these nonlinear systems in order to understand how this stability character can be affected. For this purpose, we built the bifurcation diagrams of the systems and construct the Poincar\'e sections depending on the excitation amplitude and frequency of the external forcing term.

\section{Conformal Invariant Pure Spinor Fermionic Models of Quantum Field Theory}

\subsection{The 2-D fermionic model}

The 2-D fermion-fermion interaction is described by the conformal invariant Lagrangian
\begin{equation}
\mathcal{L}=i\bar\psi  \partial \psi+\frac{g}{2}\left(\bar \psi \psi\right)^2
\end{equation}
The fermion field $\psi\left(x\right)$ has the scale dimension $\frac{1}{2}$ [8]. The equation of motion that follows from this Lagrangian is
\begin{equation}
i\sigma_\mu \partial_\mu \psi+g\left(\bar \psi \psi\right)\psi=0
\end{equation}
where $\sigma_\mu$ are  Pauli matrices and $g$  is the positive coupling constant.

\subsection{The 4-D fermionic model}

A conventional generalization of the Lagrangian (2.1) to 4-D would be based on fermion fields of scale dimension $\frac{3}{2}$  [4]
\begin{equation}
\mathcal{L}=i\bar\psi \partial \psi+g\left(\bar \psi \psi\right)^\frac{4}{3}
\end{equation}
The equation of motion that follows from this Lagrangian is
\begin{equation}
i\gamma_\mu \partial_\mu \psi+g\left(\bar \psi \psi\right)^\frac{1}{3}\psi=0
\end{equation}
where $\gamma_\mu$ are $4\times 4$ Dirac matrices, $\psi\left(x\right)$ is a four-component complex spinor function, $\bar \psi=\psi^\dagger \gamma_0$ and $g$ is the positive coupling constant. It was shown that Eq. (2.4) is invariant under conformal transformations [4]. This 4-D conformally invariant pure fermionic model with a nonlinear self-coupled spinor term is similar to the Heisenberg$'$s nonlinear generalization of Dirac$'$s equation [2].

\section{Instanton Solutions for Fermionic Models}

In Ref. [17] a special technique was introduced to find $\bar\psi\psi$ by spontaneous symmetry breaking of the full conformal group and characterizing by being invariant under the transformations of a special subgroup which turns out to reflect the final symmetry properties of the ground state of the system. This was realized with the help of an $R_\mu$ operator
\begin{equation}
R_\mu=\frac{1}{2}\left(aP_\mu+\frac{1}{a}D_\mu\right)
\end{equation}
where $P_\mu$ is the 4-momentum and $D_\mu$ is the conformal scale invariance operators in the 4-D Euclidean space-time. Here $a$ is a parameter with the dimension of length. In this way the problem has been reduced to finding the solution of the following equation
\begin{equation}
R_\mu\left(\bar\psi \psi\right)
 \equiv\frac{i}{a}\Big[\frac{a^2-x^2}{2}\partial_\mu +\left(x\partial+2d\right)x_\mu\Big ] \left(\bar \psi \psi\right) =0
\end{equation}
as 
\begin{equation}
\bar \psi \psi=\pm \frac{a}{g\left(a^2+x^2\right)}
\end{equation}
This solution has been named as instanton type solution. Then the Euclidian configuration form of Heisenberg ansatz [3]
\begin{equation}
\psi =[ix_\mu\gamma_\mu \chi\left(s\right)+\varphi\left(s\right)]c
\end{equation}
with an arbitrary spinor constant $c$  was introduced. Here $\chi\left(s\right)$ and $\varphi\left(s\right)$ are real functions of $s=x^2+t^2\left(x_1=x,x_2=t\right)$

In Ref. [9], the instanton and meron types of classical solutions were found in the case of 2-D conformal invariant Lagrangian exploiting the same symmetry arguments as for gauge and scalar fields. One of them leads to the vanishing of energy-momentum tensor and finite action [8, 9]. Vanishing of energy-momentum tensor gives this solution a chance to be a candidate for a vacuum in the quantum world, since classical solutions are interpreted as vacuum expectation values. Also the instanton and meron types of classical solutions were found in the case of 4-D conformal invariant Lagrangian and shown that they correspond to a spontaneous breaking of space-time symmetries [5, 6].  

\section{Characteristic Eigenvalues of Spinor-type Instantons}

Inserting Eq.(3.4) into Eq.(2.2) we get
\begin{subequations}
\begin{equation}
2\frac{dF\left(u\right)}{du}+\frac{1}{2}F\left(u\right)  -\alpha AB\left(F\left(u\right)^2 +G\left(u\right)^2\right)G\left(u\right)=0 
\end{equation}
\begin{equation}
2\frac{dG\left(u\right)}{du}-\frac{1}{2}G\left(u\right) +\alpha AB\left(F\left(u\right)^2 +G\left(u\right)^2\right)F\left(u\right)=0 
\end{equation}
\end{subequations}
in dimensionless form. Here $F$ and $G$ are dimensionless functions of $u$; $\alpha,A,B$ are constants [5]. By defining a new constant  this system can be written as a flow
\begin{equation}
\mathbf{f}_{2D} =\left(-\frac{1}{4} F+ \frac{1}{2}\beta[F^2+G^2]G ,\frac{1}{4}G +\frac{1}{2}\beta[F^2+G^2]F \right) 
\end{equation}
We take the divergence of $\mathbf{f}_{2D}$ and see that this flow is conserved, since $\nabla.\mathbf{f}_{2D}=0$

The Jacobian of the above system is 
\begin{equation}
\noindent\resizebox{0.9\hsize}{!}{$J_{2D}=\left(\begin{matrix}
  -\frac{1}{4}+FG\beta & -F^2\beta-\frac{1}{2}\left(F^2+G^2\right)\beta \\
 G^2\beta-\frac{1}{2}\left(F^2+G^2\right)\beta & \frac{1}{4}-FG\beta \\
 \end{matrix} \right) $}
\end{equation}
and 
\begin{equation}
det\left(J_{2D}\right)=-\frac{1}{16}+\frac{FG\beta}{2}+\frac{3F^4\beta^2}{4}+\frac{3}{2}F^2G^2\beta^2+\frac{3G^4\beta^2}{4} 
\end{equation}

The characteristic eigenvalues providing an understanding of the topological behavior around the singularity points come from $|J_{2D}-\lambda I|=0$ as 
\begin{equation}
	\noindent\resizebox{0.9\hsize}{!}{$\lambda_{2D\pm}= \pm \frac{1}{4} \sqrt{1-8\beta F G-12\beta^2 F^4-12\beta^2 G^4-24 \beta^2 F^2 G^2}$}
\end{equation}
For $\beta=0$, the eigenvalues become real; the fix point is hyperbolic. For the other values, the eigenvalues are purely imaginary. So the fix points are elliptic [18].

When we follow the same steps for Eq.(2.4) [3,4,5] we get the equations with $s=x^2_\mu=r^2+t^2 \left(x_1 \equiv x,x_2 \equiv y, x_3 \equiv z, x_4 \equiv t \right) $ given below as

\begin{subequations}
\begin{equation}
2\frac{dF\left(u\right)}{du}+\frac{3}{2}F\left(u\right)
 -\alpha \left(AB\right)^\frac{1}{3}\left(F\left(u\right)^2 +G\left(u\right)^2\right)^\frac{1}{3}G\left(u\right)=0
\end{equation}
\begin{equation}
2\frac{dG\left(u\right)}{du}-\frac{3}{2}G\left(u\right)
 +\alpha \left(AB\right)^\frac{1}{3}\left(F\left(u\right)^2 +G\left(u\right)^2\right)^\frac{1}{3}F\left(u\right)=0 
\end{equation}
\end{subequations}

This system can also be written as a flow by defining a new constant $\gamma=\alpha \left(AB\right)^\frac{1}{3}$
\begin{equation}
\mathbf{f}_{4D}=\left(-\frac{3}{4}F+\frac{1}{2}\gamma[F^2+G^2]^\frac{1}{3}G  ,\frac{3}{4}G-\frac{1}{2}\gamma[F^2+G^2]F\right) 
\end{equation}
This flow is also conserved, since $\nabla.\mathbf{f}_{4D}=0$. The Jacobian is
\begin{equation}
\noindent\resizebox{0.9\hsize}{!}{$J_{4D} =\left(\begin{matrix}
  -\frac{3}{4}+\frac{F G \gamma}{3\left(F^2+G^2\right)^\frac{2}{3}} & -F^2\beta-\frac{1}{2}\left(F^2+G^2\right)\beta 
 G^2\beta-\frac{1}{2}\left(F^2+G^2\right)\beta &   \frac{3}{4}-\frac{F G \gamma}{3\left(F^2+G^2\right)^\frac{2}{3}} 
 \end{matrix} \right) $}
\end{equation}
and
\begin{equation}
det\left(J_{4D}\right)=-\frac{9}{16}+\frac{FG\gamma}{2\left(F^2+G^2\right)^\frac{2}{3}}
+\frac{F^2\gamma^2}{6\left(F^2+G^2\right)^\frac{1}{3}}+\frac{1}{4}\left(F^2G^2\right)^\frac{2}{3} \gamma^2
\end{equation}
The characteristic eigenvalues coming from $|J_{2D}-\lambda I|=0$ are
\begin{equation}
\noindent\resizebox{0.9\hsize}{!}{$\lambda_{4D\pm}=\pm \frac{\sqrt{27-\frac{24\gamma FG}{\left(F^2+G^2\right)^\frac{2}{3}}-\frac{8\gamma^2 F^2}{\left(F^2+G^2\right)^\frac{1}{3}}-\frac{8\gamma^2 G^2}{\left(F^2+G^2\right)^\frac{1}{3}}-12\gamma^2 \left(F^2+G^2\right)^\frac{1}{3}}}{4\sqrt{3}}$}
\end{equation}

For $\gamma=0$, the eigenvalues are real. Hence the fix point is hyperbolic. For the other  values, they are purely imaginary. So the fix points are elliptic [18]. The details of these calculations can be found in Ref.18.

\section{Duffing-type stability behaviour of spinor-type instantons}

Through the following investigation firstly the Duffing-type stability of our 2-D and 4-D spinor-type equations will be checked. Therefore we firstly like to touch on Duffing oscillator equation, the famous example describing the motion of a classical particle in a double well potential [19]. Various works have been performed on this equation for many years. This is due to its being a representative of many non linear systems, with its extremely rich dynamics - many complex solutions and chaotic behaviours. 

The most general form of the Duffing oscillator equation is
\begin{equation}
\ddot{x}+\delta \dot{x} \pm \mu x +\eta x^3=\mathbf{f} Cos\left(\omega t +\phi \right)
\end{equation}
which differs from the forced and damped harmonic oscillator equation only by the nonlinear term $\eta x^3$. In the above, $\delta$ is the coefficient of viscous damping and $\mathbf{f}$ is the excitation amplitude. We are going to consider this equation with the minus sign and no forcing. Written as a flow it means 
\begin{subequations}
\begin{equation}
\dot{x}=y
\end{equation}
\begin{equation}
\dot{y}= \mu x -\eta x^3-\delta y
\end{equation}
\end{subequations}
and its characteristic phase space display is presented in Fig.1a.

Using the above equations (Eq.2.2 and Eq.2.4) belonging to our 2-D and 4-D spinor-type instantons we also formed the characteristic phase space displays. These are presented in Fig.1b and Fig.1c too.

\begin{figure}
\centering
\includegraphics[width=0.42\textwidth]{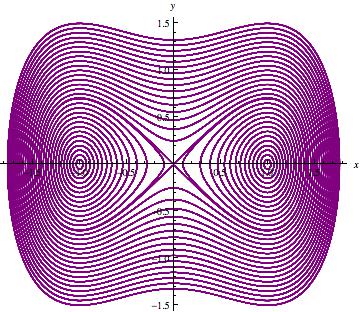} 
\includegraphics[width=0.38\textwidth]{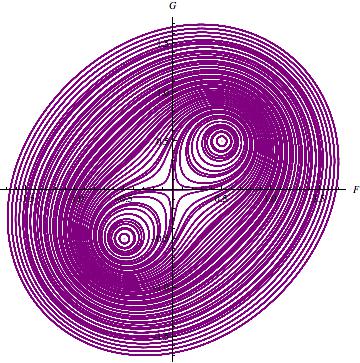} \\  (a)\hspace{4.8cm} (b) \\
\includegraphics[width=0.42\textwidth]{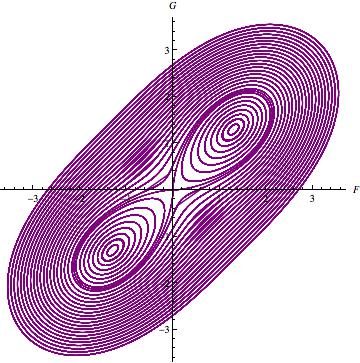} \\ (c) \\
\caption{The phase space displays of the considered (a) Duffing flow, (b) 2-D spinor wave equation (Thirring flow) and (c) 4-D spinor wave equation (Gursey flow). }
\end{figure}
As can be seen from Fig.1 (plotted for$\beta=\gamma=\mu=\eta=1$), the phase space dynamics of 2-D and 4-D spinor type instantons possess a Duffing oscillator type steady-characterization without forcing and damping [14, 20].

\section{Periodic Forcing of both Instanton Equations }

The dynamics of many nonlinear oscillators have been studied widely under sinusoidal external forcing since its easy application to real physical systems. This kind of forcing can also be helpful to control and create various nonlinear behaviours in dynamical systems. In this paper, a cosine wave will do this job.

We perturbe the 2-D fermionic model as 
\begin{subequations}
\begin{equation}
2\frac{dF(u)}{u}+\frac{1}{2}F(u)
-\beta\left(F(u)^2+G(u)^2\right)G(u)=0
\end{equation}
\begin{equation}
2\frac{dG(u)}{u}-\frac{1}{2}G(u) 
 +\beta\left(F(u)^2+G(u)^2\right)F(u)   
 =A_{2D} Cos(\omega_{2D} u)
\end{equation}
\end{subequations}
Here $A_{2D}$ is the amplitude of the external forcing and  is its frequency. This system can be converted to the following flow
\begin{subequations}
\begin{equation}
2\frac{dF(u)}{u}+\frac{1}{2}F(u) 
-\beta\left(F(u)^2+G(u)^2\right)G(u)=0 
\end{equation}
\begin{equation}
2\frac{dG(u)}{u}-\frac{1}{2}G(u)
 +\beta\left(F(u)^2+G(u)^2\right)F(u)
=A_{2D} Cos(\omega_{2D} H(u))
\end{equation}
\begin{equation}
\frac{H(u)}{du}=\Omega
\end{equation}
\end{subequations}
$\Omega$ adds an extra dimension for the numerical calculations. The vector field is
\begin{equation}
\mathbf{f}_{F2D}=(-\frac{1}{4}F+\frac{1}{2}\beta[F^2+G^2]G
 ,\frac{1}{4}G-\frac{1}{2}\beta[F^2+G^2]F+A_{2D} Cos(\omega_{2D} H), \Omega)
\end{equation}
and its divergence $\nabla.\mathbf{f}_{F2D}=0$ so the flow is conservative under the external forcing. In a conservative system a collection of trajectories will occupy a constant volume of a region of phase space as the trajectories move around under the dynamics of the system. This is the famous Liouville$'$s Theorem [21].

We also perturbe the 4-D fermionic model in the same way. 
\begin{subequations}
\begin{equation}
2\frac{dF(u)}{u}+\frac{3}{2}F(u)
-\gamma\left(F(u)^2+G(u)^2\right)^\frac{1}{3}G(u)=0
\end{equation}
\begin{equation}
2\frac{dG(u)}{u}-\frac{3}{2}G(u)
+\gamma\left(F(u)^2+G(u)^2\right)F(u) 
=A_{4D} Cos(\omega_{4D} u)
\end{equation}
\end{subequations}
and can be converted to the flow
\begin{subequations}
\begin{equation}
2\frac{dF(u)}{u}+\frac{3}{2}F(u)
-\gamma\left(F(u)^2+G(u)^2\right)^\frac{1}{3}G(u)=0
\end{equation}
\begin{equation}
2\frac{dG(u)}{u}-\frac{3}{2}G(u)
+\gamma\left(F(u)^2+G(u)^2\right)^\frac{1}{3}F(u)
=A_{4D} Cos(\omega_{4D} K(u))
\end{equation}
\begin{equation}
\frac{K(u)}{du}=\Gamma
\end{equation}
\end{subequations}

The constant $\Gamma$ adds an extra dimension for numerical calculations. The vector field is
\begin{equation}
\mathbf{f}_{F4D} =(-\frac{3}{4}F+\frac{1}{2}\gamma[F^2+G^2]^\frac{1}{3}G
,\frac{3}{4}G-\frac{1}{2}\gamma[F^2+G^2]^\frac{1}{3}F+A_{4D} Cos(\omega_{4D} K), \Omega )
\end{equation}
and has a vanishing divergence $\nabla.\mathbf{f}_{F4D}=0$ so the flow is conservative.

\section{Numerical Result}

We demonstrated above that the systems are conservative under external forcing. Now it is well known that chaotic behaviour is not an exclusive property of dissipative systems but of conservative systems as well. However the phenomenology of chaos in conservative systems is somewhat different from that in the dissipative systems. As opposed to a conservative system, the initial volume in the phase space is not conserved during the evolution in dissipative systems. A conservative system can move along any available trajectory, depending on its energy [22]. Studying chaos in conservative systems helps illuminate many important properties of classical nonlinear dynamics and emphasize some important problems in quantum-classical correspondence. Also checking the Duffing type stability of spinor type instantons against external forcing may provide us with some insight on the subject. To that end, we are going to look for the dynamical behaviours of spinor type instantons by changing the forcing parameters numerically by constructing bifurcation diagrams and Poincar\'e maps.

\begin{figure}
\centering
\includegraphics[width=0.4\textwidth]{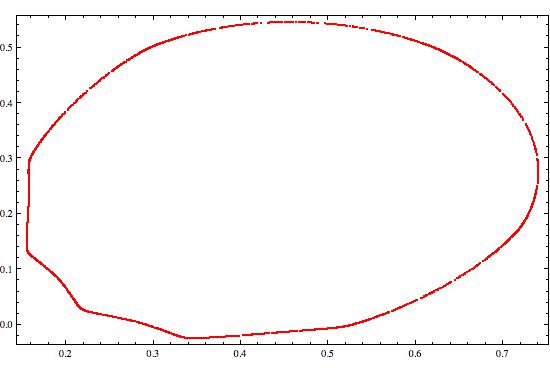} \hspace{0.5cm}
\includegraphics[width=0.415\textwidth]{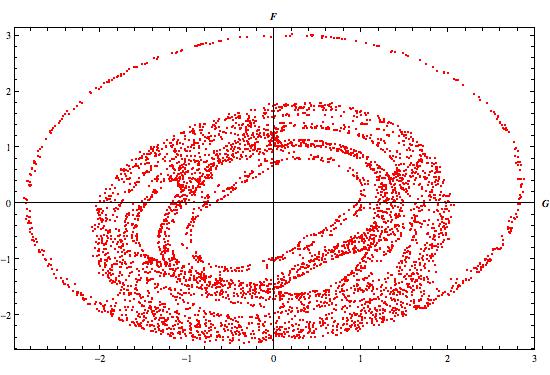} \\ (a)\hspace{5.0cm}  (b)  \\
\includegraphics[width=0.4\textwidth]{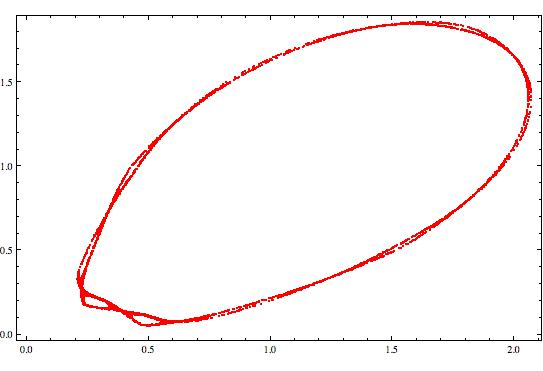} \hspace{0.5cm}
\includegraphics[width=0.415\textwidth]{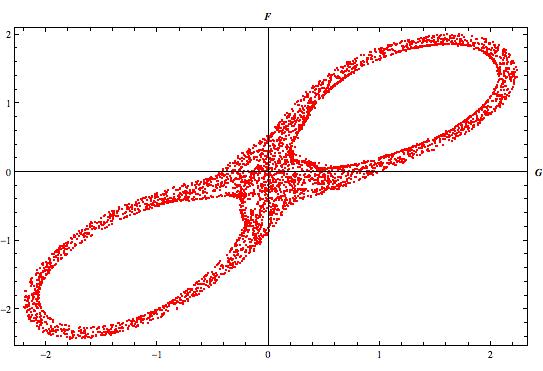} \\ (c)\hspace{5.0cm}  (d)  \\
\caption{Sensitive dependence on initial conditions at $\omega=\pi, A=1$ for 2-D spinor wave equation; (a) $F(0)=0.7, G(0)=0.25,H(0)=0$; (b) $F(0)=0.71, G(0)=0.251,H(0)=0$ and 4-D spinor wave equation at $\omega=\frac{\pi}{2}, A=1$ (c) $F(0)=0.47, G(0)=0.97,K(0)=0$ (d) $F(0)=0.471, G(0)=0.975,K(0)=0$. }
\end{figure}
It is well known that chaotic systems sensitively depend on the initial conditions and the transitions from regular states to chaos are caused by insignificant changes in the initial conditions. To see this extreme sensitivity of spinor type instantons to initial conditions with excitation, in Fig.2, we show the Poincar\'e sections corresponding to regular and chaotic behaviours with two different very close initial conditions. For the 2-D system, as is seen from Fig. 2(a), the flow is a closed orbit thus the behaviour is regular for $F(0)=0.7, G(0)=0.25$ and $H(0)=0$ at $\omega=\pi$ and $A=1$, in fact a quasiperiodic one. If we take another initial condition which is very close to the first one $\left(F(0)=0.71, G(0)=0.251,H(0)=0\right)$ we observe chaotic orbits in Fig. 2(b). Alike, the sensitive dependence on initial conditions of 4-D system is shown in Figures 2(c) and (d). 

\begin{figure}
\centering
\includegraphics[width=0.47\textwidth]{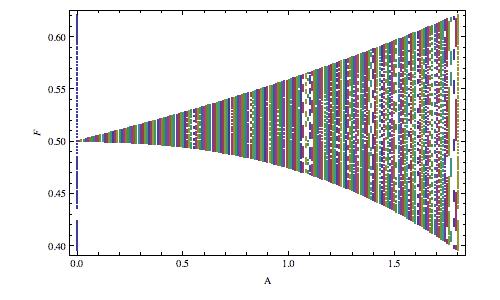}
\includegraphics[width=0.47\textwidth]{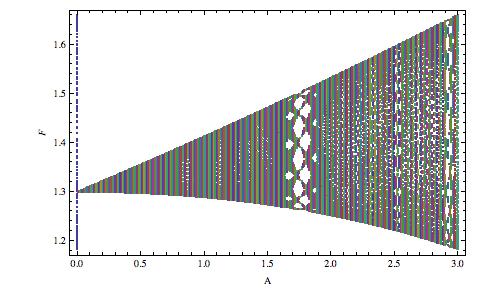}\\ (a)\hspace{5.1cm}  (b)  \\
\caption{Bifurcation diagrams in ( $A$,$F$ )-plane for (a) 2-D model at $\omega=\pi$; (b) 4-D model at $\omega=\frac{\pi}{2}$ }
\end{figure}
Bifurcation means fundamental change in the nature of a solution and the bifurcation diagram provides us a useful way to show how a nonlinear system$'$s behaviour changes with the control parameter [23]. In the forced system one can consider two main control parameters: the amplitude and the frequency of excitation. The corresponding bifurcation diagrams of 2-D and 4-D spinor type instantons in the $\left(A ,F\right)$-plane are shown in Fig. 3, respectively. A lot of confused points falling on the diagrams correspond to a chaotic behaviour [24]. If we analyze the bifurcation diagrams of the systems, we observe the windows of order, which the systems briefly leave their chaotic states and then rapidly return to chaos as   increases. These short intervals are quite periodic or quasi-periodic.

\begin{figure}[t]
\centering
\includegraphics[width=0.40\textwidth]{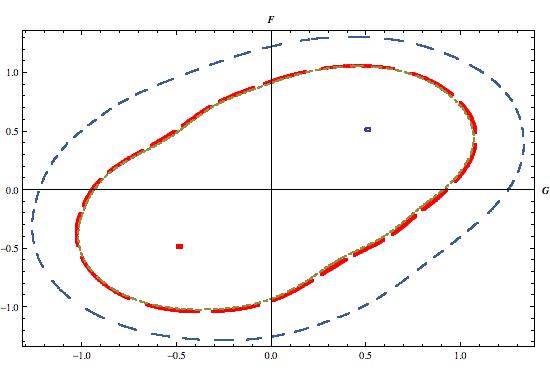}\hspace{0.5cm}
\includegraphics[width=0.40\textwidth]{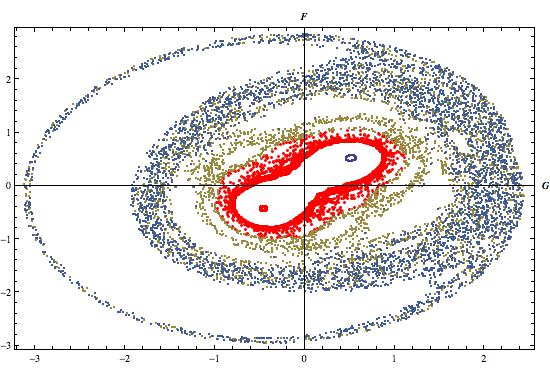}\\ (a)\hspace{5.0cm}  (b)  \\
\includegraphics[width=0.40\textwidth]{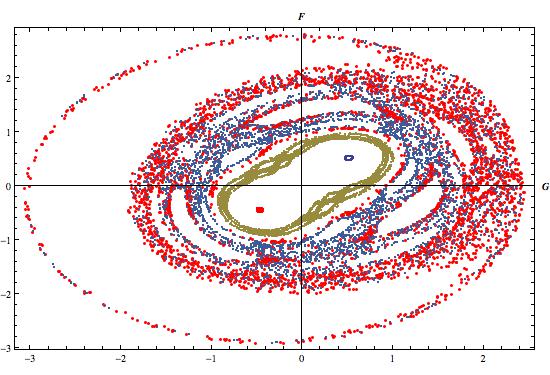}\hspace{0.5cm}
\includegraphics[width=0.40\textwidth]{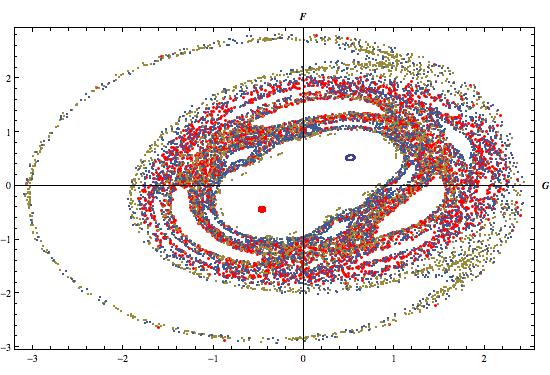}\\ (c)\hspace{5.0cm}  (d)  \\
\caption{Poincar\'e sections of 2-D spinor wave equation at $\omega=\pi$   for different initial conditions (a) $A=0.57$  , (b) $A=0.943$ , (c) $A=1.017$  and (d) $A=1.27$ .}
\end{figure}
We illustrate in Fig.4 the regular and chaotic behaviours of 2-D system for some random possible initial values keeping $\omega=\pi$ and $\beta=1$ for  $A=0.57$  , $A=0.943$, $A=1.017$  and $A=1.27$, respectively. It is interesting that the obtained phase space displays are typical for Kolmogorov-Arnold-Moser (KAM) like dynamics. It is well known that the phase spaces having a KAM-like structure are fundamentally different from the controlling dissipative chaotic attractors. Some originally periodic solutions remain regular and mean quasi periodicity while others behave chaotically [25]. For the weak driving in Fig. 4(a) the system shows regular behaviour in harmony with the bifurcation diagram. As we reinforce the driving, Fig. 4(b)-4(d) shows that the chaotic orbits appear in the region near the centre of phase space. As is seen from Fig.1 (b) the two-dimensional spinor type instantons lies on the separable closed curves that correspond to regular trajectories. But the external force destroys the seperatrix and a number of chaotic instantons appears depending on the excitation amplitude of the external forcing.

\begin{figure}[t]
\centering
\includegraphics[width=0.40\textwidth]{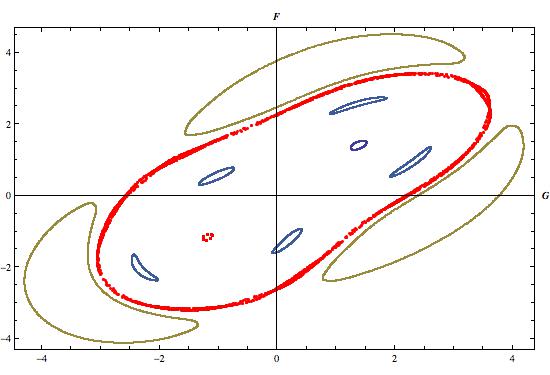}\hspace{0.5cm}
\includegraphics[width=0.40\textwidth]{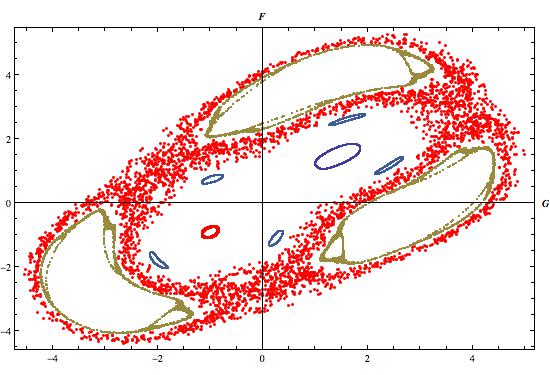}\\ (a)\hspace{5.0cm}  (b)  \\
\includegraphics[width=0.40\textwidth]{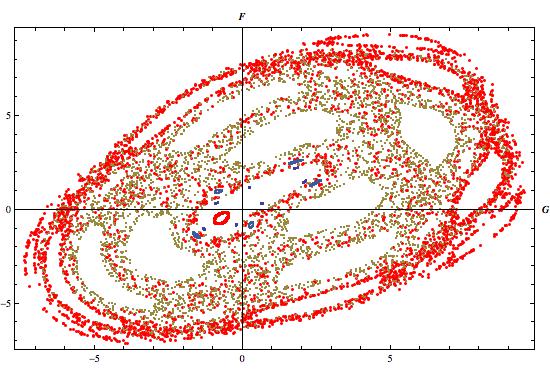}\hspace{0.5cm}
\includegraphics[width=0.40\textwidth]{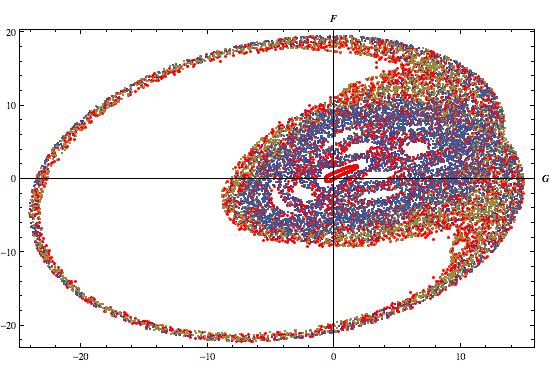}\\ (c)\hspace{5.0cm}  (d)  \\
\caption{Poincar\'e sections of 4-D spinor wave equation at $\omega=\frac{\pi}{2}$   for different initial conditions (a) $A=0.87$  , (b) $A=0.91$ , (c) $A=1$  and (d) $A=1.215$ .}
\end{figure}
Alike, 4-D model$'$s phase space displays possess a KAM-like structure that can easily be seen from Fig.5. A number of chaotic instantons appears in a 4-D conformally invariant pure spinor model with a nonlinear self-coupled spinor term depending on the values of the excitation amplitude of the external forcing. The obtained results indicate that 2-D and 4-D pure fermion models with fractional nonlinearity have a rich set of excited fermion states (in harmony with the results in Ref. [26]). Nonlinearity occurs such as the jump phenomenon, i.e. the stable behaviour changes considerably due to a transition from one stable solution to another solution as the excitation amplitude is varied with the control parameter.

\section{Conclusion}

It is well known that instantons are considered as configurations of quantum fields that provide tunnelling effect between different topology vacua in space-time. Tunnelling plays an important role in gauge field theories. As is known classical gauge field theories are inherently chaotic [27]. Therefore the study of chaos in instanton physics based on chaos criterion in quantum field theory is also of required interest.
In this paper, firstly we show the sensitive dependence on the initial conditions of excited spinor-type instantons. The transitions from regular states to chaos are caused by insignificant changes in the initial conditions for both pure spinor fermionic models. Secondly, the dynamics of spinor-type instantons are studied in two and four dimensions driven by the cosine wave by constructing the bifurcation diagrams and Poincar\'e sections. We notice the similarities in the bifurcation structures of both systems in the presence of periodic force. We observe the periodic windows that the systems briefly leave their chaotic states and then rapidly return to chaos. As is seen from Fig.1, the 2-D and 4-D spinor type instantons lies on the separable closed curves that correspond to regular trajectories. The obtained results from Poincar\'e sections and bifurcation diagrams show vanishing of Duffing-type stability characteristics of spinor-type instantons in phase space depending on the external forcing parameter values. The external force destroys the seperatrix and forms a stochastic layer. In this layer, a number of chaotic instantons appears for both systems similar to each other. There are different types of studies about chaotic instantons [28-31]; but this is the first observation of common behaviours of chaotic instantons in spinor-type fermionic models. Also phase space of the flow for both systems possesses a KAM-like structure, i.e. some originally periodic solutions remain regular and mean quasi-periodicity while others behave chaotically [25].The observed regular and chaotic behaviours of both excited systems in Poincar\'e sections for various deriving force amplitude values are in harmony with the bifurcation diagrams.
In the view of the obtained results, we can interestingly conclude that although models have different quantum spinor numbers and dimensions, the behaviours of spinor-type instantons in 4-D fermionic model are similar to those of spinor-type instantons in 2-D fermionic model under the effect of cosine wave forcing. Since the knowledge of the critical behaviour of a conformal symmetric quantum field theory near the fixed points, this result is not surprising [32-36]. The distinguishing feature of our work is of being the first remark of the spinor-type chaotic instantons in phase space not dependent on the quantum fractional spinor number as well as dimensions.

\section*{Acknowledgements}
This work was supported by the Scientific Research Projects Coordination Unit of Istanbul University; Project no: 35737.

\address{Department of Physics, Istanbul University \\Istanbul, Turkey \\
\email{fatma.aydogmus@gmail.com; fatmaa@istanbul.edu.tr}}

\end{document}